\documentclass[twocolumn,showpacs,amsmath,amssymb,prl]{revtex4}
\usepackage{graphicx}
\usepackage{dcolumn}
\usepackage{bm}
\usepackage{psfig}

\newcommand{\ppbar}{p \overline{p}}
\newcommand{\pbar}{\overline{p}}
\newcommand{\Lambar}{\overline{\Lambda}}
\newcommand{\Sigbar}{\overline{\Sigma^0}}
\newcommand{\jpsi}{J/\psi}

\begin{document}
\preprint{Draft-PRL}

\title{ \bf \boldmath Observation of a threshold enhancement 
in the $p \Lambar$ invariant mass spectrum}

\author{M.~Ablikim$^{1}$, J.~Z.~Bai$^{1}$, Y.~Ban$^{10}$, J.~G.~Bian$^{1}$, 
X.~Cai$^{1}$, J.~F.~Chang$^{1}$, H.~F.~Chen$^{16}$, H.~S.~Chen$^{1}$, 
H.~X.~Chen$^{1}$, J.~C.~Chen$^{1}$
% ...
\vspace{0.2cm} \\(BES Collaboration)\\
\vspace{0.2cm}
$^1$ Institute of High Energy Physics, Beijing 100039, People's Republic of China\\
% ...
$^{10}$ Peking University, Beijing 100871, People's Republic of China\\
% ...
$^{16}$ University of Science and Technology of China, Hefei 230026, People's Republic of China\\
% ...
}

\date{\today}

\begin{abstract}
{ An enhancement near the $m_p + M_{\Lambda}$ mass threshold is observed in 
the combined $p\Lambar$ and $\pbar\Lambda$ invariant mass spectrum 
from $J/\psi \rightarrow p K^- \Lambar + c.c.$ decays. It can be fit 
with an S-wave Breit-Wigner resonance with a mass 
$m=2075\pm 12 \:({\rm stat}) \pm 5 \:({\rm syst})$~MeV and a width of
$\Gamma =90 \pm 35 \:({\rm stat}) \pm 9 \:({\rm syst})$~MeV; it can also be
fit with a P-wave Breit-Wigner resonance.  Evidence for a similar enhancement 
is also observed in $ \psi' \rightarrow p K^- \Lambar + c.c.$ decays.  
The analysis is based on samples of $5.8 \times 10^7$ $J/\psi$ and 
$1.4 \times 10^7$ $\psi'$ decays accumulated in the BES II detector at 
the Beijing Electron-Positron Collider.  }
\end{abstract}
\maketitle


An anomalous enhancement near the mass threshold in the 
$p\pbar$ invariant mass spectrum was observed by the BES II experiment in 
$\jpsi\rightarrow\gamma p\pbar$ decays ~\cite{gpp}. This enhancement 
can be fit with an S-wave Breit-Wigner resonance function with a mass 
around 1860~MeV and a width $\Gamma<30$~MeV, and has been interpreted 
as a possible baryonium state \cite{baryonium}. Similar  $p\pbar$ 
mass-threshold enhancements have been observed in the decays 
$B^+ \rightarrow K^+ \ppbar$ and $\bar{B}^0 \rightarrow D^0 \ppbar$  
by the Belle Collaboration \cite{belle_pp1,belle_pp2}. These somewhat 
surprising experimental observations have stimulated a number of theoretical 
speculations \cite{baryonium,theory}. It is, therefore, of special interest 
to search for possible resonant structures in other baryon-antibaryon final 
states. The Belle Collaboration recently observed a near-threshold enhancement 
in  the $p\Lambar$ mass spectrum from $B \rightarrow p\Lambar\pi$ 
decays \cite{belle_pL}. In this letter, we report the observation of
an enhancement near threshold in the $p \Lambar$ invariant mass spectrum in 
$J/\psi \rightarrow p K^- \Lambar$ and in $ \psi' \rightarrow p K^- \Lambar$ 
decays. (In this letter the inclusion of charge conjugate modes is always 
implied). The results are based on an analysis of $5.8 \times 10^7$ $J/\psi$ 
and $1.4 \times 10^7$ $\psi '$ decays detected  in the upgraded Beijing 
Spectrometer (BESII) at the Beijing Electron-Positron Collider (BEPC).


BESII is a large solid-angle magnetic spectrometer that is described
in detail in Ref.\cite{BESII}. Charged particle momenta are determined with a
resolution of $\sigma_p/p = 1.78\%\sqrt{1+p^2(\mbox{\rm GeV}^2)}$ in a 
40-layer cylindrical main drift chamber (MDC). Particle identification is 
accomplished by specific ionization ($dE/dx$) measurements in the MDC and 
time-of-flight (TOF) measurements in a barrel-like array of 48 scintillation 
counters. The $dE/dx$ resolution is $\sigma_{dE/dx} = 8.0\%$; the TOF 
resolution is measured to be $\sigma_{TOF} = 180$~ps for Bhabha events. 
Outside of the time-of-flight counters is a 12-radiation-length barrel shower 
counter (BSC) comprised of gas tubes interleaved with lead sheets. 
The BSC measures the energies and directions of photons with resolutions of
$\sigma_E/E\simeq 21\%/\sqrt{E(\mbox{GeV})}$, $\sigma_{\phi} = 7.9$ mrad, 
and $\sigma_{z}$ = 2.3 cm. The iron flux return of the magnet is instrumented 
with three double layers of counters that are used to identify muons. In this 
analysis, a GEANT3-based Monte Carlo (MC) package with detailed consideration 
of the detector performance (such as dead electronic channels) is used. The 
consistency between data and MC has been carefully checked in many high purity
physics channels, and the agreement is reasonable.

\begin{figure}[hbtp]
          \centerline{\psfig{file=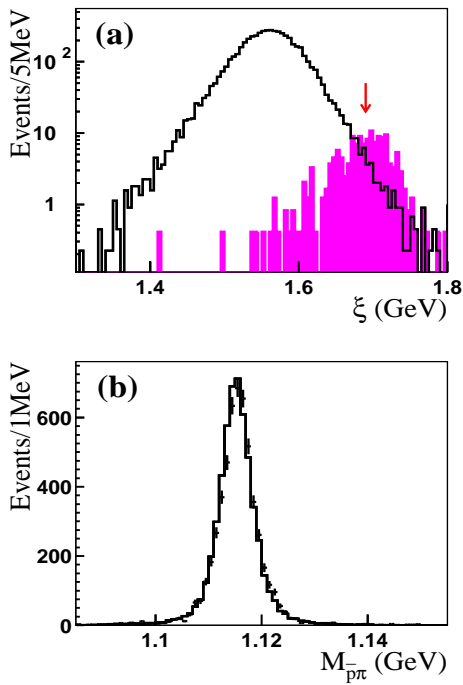,height=9.0cm,width=6.0cm}}
          \caption{ 
           (a) The $\xi$ distribution (see text). The solid
               histogram denotes $\jpsi \rightarrow pK\Lambar$ 
               events, and the shaded histogram 
               $\jpsi \rightarrow pK\Sigbar$ events, where both histograms 
               are normalized to $5.8 \times 10^7 \jpsi$ events.  
               Events with $\xi$ values below the arrow are selected.
	   (b) The $\pbar\pi$ invariant mass distribution for selected
               events; points with error bars denote the data and 
               the histogram the MC (normalized to data). }
	\label{cutsig}
	\end{figure}

The $J/\psi \rightarrow pK^-\Lambar$ candidate events are required to 
have four charged tracks, each of which is well fitted to a helix within the 
polar angle region $|\cos \theta|<0.8$ and with a transverse momentum larger 
than $50$~MeV. The total charge of the four tracks is required to be zero.
For each track, the TOF and $dE/dx$ information are combined to form particle 
identification confidence levels for the $\pi, K$ and $p$ hypotheses; the 
particle type of a track is assigned to be that of the hypothesis with the 
largest confidence level. In this analysis, reliable identification of the 
$K^-$ is important. To have high efficiency, it is only required that one 
track be positively identified as a proton or antiproton. Events where the 
$p, ~ K^- , ~ \pbar $ and $ ~\pi^+ $ tracks are all unambiguously identified
are subjected to a four-constraint (4C) kinematic fit with the corresponding 
mass assignments for each track. For events with ambiguous particle 
identification, all possible 4C combinations are formed, and the combination 
with the smallest $\chi^2$ is chosen. The final $\chi^2$ is required to be 
less than 20.  Further, the $p$ and $K^-$ tracks are required to originate 
near the interaction point, and the invariant mass of the $\pbar \pi^+$ 
combination is required to be less than $1.15$~GeV. To suppress background
events from $\jpsi\rightarrow pK^-\Sigbar$, we require
$ \xi = E_{miss}+1.39M_{pK~miss} < 1.69$~GeV (see Fig. \ref{cutsig}(a)), 
where $E_{miss}$ denotes the difference between the center-of-mass energy 
(3.097 GeV) and the total energy of the four charged tracks, and $M_{pK~miss}$
denotes the mass recoiling against the proton-kaon system. This selection 
criterion is determined by optimizing the signal to background ratio based on 
Monte Carlo simulations. A sample of  $5421$ $J/\psi \rightarrow pK^-\Lambar$ 
candidates survive the final selection. The $\pbar\pi^+$ invariant mass 
spectrum for these events, where a clear $\Lambar\rightarrow \pbar\pi^+$ 
signal is evident, is shown in Fig. \ref{cutsig}(b).

The $J/\psi \rightarrow pK^-\Lambar$ events are experimentally quite distinct:
they contain only charged tracks, three of which are heavy particles
(i.e., $p, \pbar$ and $K$), and the kinematics strongly constrains the event 
selection and mass assignments. In order to maintain a high selection 
efficiency and reduce systematic uncertainties, positive identification of 
only the proton or antiproton is required, and no requirement is placed on the
$\Lambda$'s secondary vertex.  The clean $\Lambar$ signal in the $\pbar\pi^+$ 
invariant mass spectrum and good agreement with the $\Lambar$ signal from a 
MC sample of $J/\psi \rightarrow pK^-\Lambar$ events (Fig. \ref{cutsig}(b)) 
indicate that the purity of the selected events is very high.


The level of background in the selected event sample was determined with two 
different MC studies.  One used a specific set of background processes: 
$\jpsi \rightarrow p K^-\Sigbar$; $\Lambda \Lambar$; $\Lambda\Lambar \pi^0$; 
$\ppbar \pi^+\pi^-$; $\ppbar \pi^+\pi^- \pi^0$; and $\Sigma^0\Sigbar$, all 
produced according to branching ratios from Particle Data Group (PDG) Tables 
\cite{pdg}.  The fraction of these events that survive the 
$J/\psi \rightarrow pK^-\Lambar$ selection criteria corresponds to about 18 
events in the selected data sample. The second study used an inclusive MC 
sample of 30 million $J/\psi$ events generated according to the LUND model 
\cite{chenjc}.  This study predicts that there are 56 background events 
in the data sample.  These studies indicate that the background in the 
selected event sample is at the $1 \sim 2\%$ level.

\begin{figure}[hbtp]
        \centerline{\psfig{file=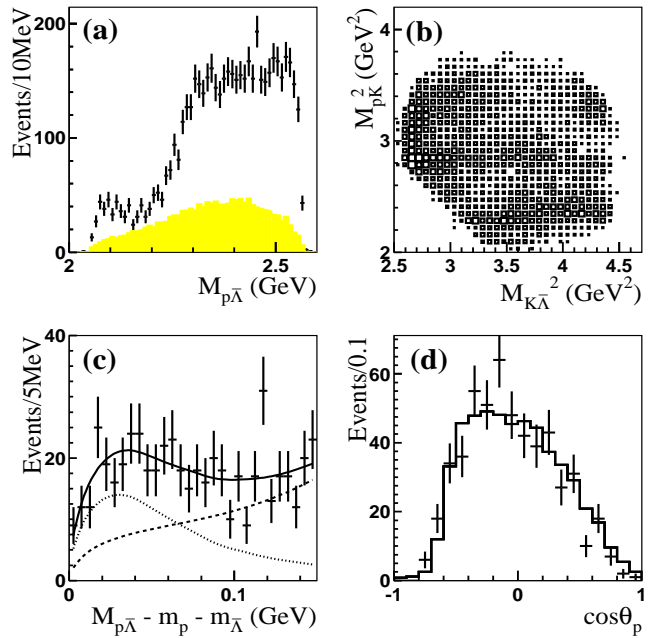,height=8.4cm,width=8.4cm}}
	\caption{ 
	    (a) The points with error bars indicate the measured 
            $p\Lambar$ mass spectrum; the shaded histogram indicates  
            phase space MC events (arbitrary normalization).
	    (b) The Dalitz plot for the selected event sample.
	    (c) A fit (solid line) to the data. The 
	    dotted curve indicates the Breit-Wigner signal and the 
            dashed curve the phase space `background'.
	    (d) The $\cos \theta_p$ distribution under the enhancement,
                the points are data and the histogram is the MC
              (normalized to data)}
	\label{x208}
	\end{figure}

The $p\Lambar$ invariant mass spectrum for the selected events is shown 
in Fig.~\ref{x208}(a), where an enhancement is evident near the mass 
threshold.  No corresponding structure is seen in a sample of
$J/\psi \rightarrow p K^- \Lambar$ MC events generated with a uniform 
phase space distribution.  The $pK^- \Lambar$ Dalitz plot is shown in 
Fig.~\ref{x208}(b).  In addition to bands for the well established 
$\Lambda^*(1520)$ and $\Lambda^*(1690)$, there is a significant $N^*$ band 
near the $K^-\Lambar$ mass threshold, and a $p\Lambar$ mass 
enhancement, isolated from the $\Lambda^*$ and $N^*$ bands, in the 
right-upper part of the Dalitz plot.

This enhancement can be fit with an acceptance weighted S-wave Breit-Wigner 
function~\cite{bw}, together with a function $f_{PS}(\delta)$ describing 
the phase space contribution, as shown in Fig.~\ref{x208}(c), 
where $f_{PS}(\delta)=N(\delta^{1/2}+a_1\delta^{3/2}+a_2\delta^{5/2})$,
$\delta=m_{p\Lambar}-m_p-m_{\Lambar}$, and the parameters $a_1$ and $a_2$ are 
determined from a fit to the $p K^- \Lambar$ MC sample events generated 
with a uniform phase-space distribution. The fit is confined to the
$ M_{p \Lambar} - M_p - M_{\Lambar} < 150$~MeV mass region and 
gives a peak mass of $m=2075\pm 12$~MeV and a width $\Gamma=90 \pm 35$~MeV.  
The fit confidence level is $22.5\% $( $\chi^2/d.o.f. = 31.1/26$), and 
$-2 ln L = 29.9$. The no resonance hypothesis is also tested, and the fit is 
much poorer: the confidence level is $5.5 \times 10^{-10}$ 
($\chi^2 /d.o.f. = 101.5/29$), and $-2 ln L = 96.2$. This indicates that the 
enhancement deviates from the shape of the phase space contribution with a 
statistical significance of about $ 7\sigma$. 

The fit yields $N_{res}=238 \pm 57$ signal events, corresponding to
a branching ratio
\begin{eqnarray}
& & BR(J/\psi \rightarrow K^-X) BR (X\rightarrow p \Lambar) 
\nonumber  \\
& = & \frac{N_{res}/(2\epsilon BR(\Lambda\rightarrow p
\pi))}{N_{J/\psi}}=(5.9\pm 1.4)\times 10^{-5},  \nonumber 
\end{eqnarray}
where $BR(\Lambda\rightarrow p \pi ) = 63.9 \pm 0.5 \%$ is taken from
the PDG, $N_{J/\psi}=(5.77 \pm 0.27)\times10^7$ is the total number of 
$J/\psi$ events \cite{ssfang}, and $\epsilon=5.47\pm 0.05\%$ is the 
MC-determined signal acceptance. 

The signal acceptance and the phase space shape $f_{PS}(\delta)$ are corrected
for differences between the low momentum $p$ and $\pbar$ tracking 
efficiencies for MC and data.  The $p$ and $\pbar$ tracking efficiencies
are measured with the data using a sample of 
$\jpsi \rightarrow \ppbar \pi^+\pi^-$ events. 

A P-wave Breit-Wigner signal function (angular momentum $L=1$) also gives an 
adequate fit to the data; here $\chi^2/{d.o.f.} = 32.5/26$ with a mass 
$M=2044 \pm 17$~MeV and a width $\Gamma = 20 \pm 45$~MeV, which is consistent 
with zero.  Fits with higher angular momentum hypotheses $L \ge 2$ fail; 
such states are expected to be strongly suppressed near threshold.

The low acceptance for low momentum protons and anti-protons produces
a non-uniform acceptance across the $M_{p\Lambar} = 2075$~MeV band in 
the Dalitz plot (Fig.~\ref{x208}(b)).  This is reflected in the non-uniform 
$\cos\theta_p$ distribution, where $\theta_p$ is the decay angle of p in the 
$p\Lambar$ CM frame, for the events in the enhancement region 
($ M_{p\Lambar} - M_p - M_{\Lambar} < 150$~MeV), as shown in 
Fig.~\ref{x208}(d). The distribution agrees well with that of a MC sample of 
$J/\psi\rightarrow K X \rightarrow K p \Lambar$ with $M_X = 2075$~MeV 
and $\Gamma_X = 90$~MeV.  Since the MC $\cos\theta_p$ distribution is 
generated as a uniform S-wave distribution, but the detected MC distribution 
agrees with data in Fig.~\ref{x208}(d), the observed distribution for the 
enhancement is consistent with S-wave decays to $p\Lambar$.

\begin{figure}[hpbt]
        \centerline{\psfig{file=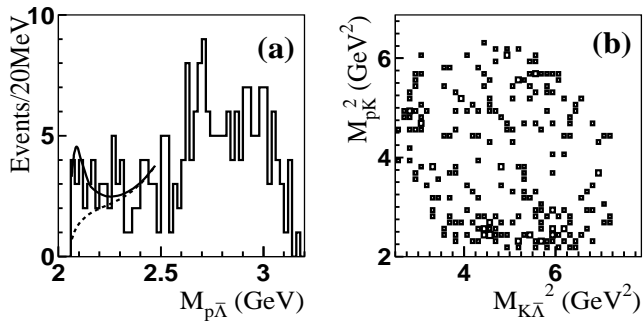,height=4.2cm,width=8.4cm}}
        \caption{Results for $\psi^\prime \rightarrow p K^- \Lambar$ 
	events: (a) A fit (solid line) to the data sample (histogram); 
	    the dashed line indicates the
	    phase space `background' contribution.
	 (b)The Dalitz plot.}
        \label{kxprime}
        \end{figure}

Evidence of a similar enhancement is observed in 
$ \psi^{'} \rightarrow p K^- \Lambar$, shown in Fig.~\ref{kxprime} (a), 
when the same analysis is performed on the $\psi^\prime$ data sample.
A fit is applied on the $\psi^\prime$ data sample with the $X(2075)$ 
parameters fixed at the values obtained from the $J/\psi$ data, i.e., 
$M_X=2075$~MeV and $\Gamma_X = 90$~MeV. The fit shows that the threshold 
enhancement in $\psi^\prime$ data  deviates from the shape of the phase space 
contribution with a statistical significance of about $4.0 \sigma$, where the 
significance is estimated from a comparison of log-likelihood values of the 
fits with and without the X(2075) signal function. 

The possibility that the enhancement in the $\jpsi$ data sample is due
to interference between $N^*$'s and $\Lambda^*$'s has been investigated with 
a partial wave analysis (PWA). The PWA results show that if the enhancement 
were from a pure interference effect, many large branching ratio $\jpsi$ 
decays to $N^*$'s and $\Lambda^*$'s near the kinematic threshold are required,
along with large mutual destructive interferences that cancel these large 
production rates \cite{pwa}.  Also, the similar enhancements seen in the 
$\psi'$ data sample and in $B \rightarrow p\Lambar\pi$ observed by the 
Belle experiment cannot be due to $N^*$ and $\Lambda^*$ interference effects
since in these cases, contributions of the signal are far from the $N^*$ and 
$\Lambda^*$ bands in the Dalitz plot (See Fig.~\ref{kxprime} (b)).

Systematic uncertainties from different sources are studied. In the above fit,
the phase space contribution is treated as the `background' under the 
enhancement. Alternative `background' shape parameters, including $N^*$'s and 
$\Lambda^*$'s contribution obtained from PWA fits, are used to estimate 
systematic uncertainties from the `background' shape. The fitting bias near 
threshold is checked by MC studies. A set of MC samples combining a signal 
(resonance near threshold) process with a uniform phase space process are 
generated. In each MC sample, the mass, width and number of signal events are 
obtained from a fit using the same procedure as that done on the data. The 
averaged offsets between the fit output values and their input values are 
taken as one source of systematic uncertainty (fitting bias). The systematic 
uncertainty from the tracking efficiencies, especially from the low momentum 
$p$ and $\pbar$ tracks, are checked from data and MC comparisons, where the 
tracking efficiencies of $p$ and $\bar p$ are determined from a data sample of
$\jpsi \rightarrow \ppbar \pi^+\pi^-$ events, and the tracking efficiencies 
of charged pions is obtained from $\jpsi \rightarrow \Lambda \Lambar$, 
$\rho \pi$ events. The systematic uncertainty from the kinematic fit is 
estimated by using a different MDC wire resolution simulation model. 
Systematic uncertainties from other sources (such as mass resolution) are 
negligible. The systematic uncertainties determined from the above studies 
are listed in Table \ref{error}, and the total systematic errors on the mass, 
width and branching ratios are 5~MeV, 9~MeV and $33.5\%$ respectively.

\begin{table}
\caption{Systematic Errors}
\centering
\begin{tabular}{|c|c|c|c|}
    \hline
~  &mass($$MeV)& width(MeV) &      BR ($\%$) \\ \hline
  `Background' shape    & 4  & 8  & 27.3  \\ 
  Fitting bias          & 3 & 3  & 12.6  \\ 
  Particle identification  & \    & \     & 3.5  \\ 
  Tracking efficiency   & 0.3  & 1.2   & $12.6$  \\ 
  $\chi^2$               & \  & \   & 5.3   \\ 
  $N_{J/\psi}$           & \    & \     & 4.7  \\ \hline
  Total                  & 5 & 9  & 33.5 \\ \hline
\end{tabular}
\label{error}
\end{table}


In summary, an anomalous enhancement near threshold is observed in the 
invariant mass spectrum of $p\Lambar$ in the 
$\jpsi \rightarrow pK^-\Lambar$ and $\psi' \rightarrow pK^-\Lambar$ processes.
Both S-wave and P-wave Breit-Wigner resonance functions can fit the 
enhancement. If it is fitted with an S-wave Breit-Wigner resonance function, 
the mass is $m=2075\pm 12 \pm 5$~MeV, the width is 
$\Gamma =90 \pm 35 \pm 9$~MeV, and the branching ratio is 
$BR(J/\psi \rightarrow K^-X) BR (X\rightarrow p\Lambar)$
= $ (5.9 \pm 1.4 \pm 2.0 ) \times 10^{-5} $, where the first errors are 
statistical and the second are systematic. To understand the nature of this 
anomalous enhancement, searching for the same enhancement in $K\pi$ and 
$K\pi\pi$ modes in the $\jpsi, \psi' \rightarrow KK\pi, KK\pi\pi$ processes
would help to distinguish whether it is from a conventional $K^*$ meson or 
from a possible multi-quark state.

The BES collaboration acknowledges the staff of BEPC for their hard efforts. 
This work is supported in part by the National Natural Science Foundation
of China under contracts Nos. 19991480, 10225524, 10225525, the Chinese
Academy of Sciences under contract No. KJ 95T-03, the 100 Talents Program of 
CAS under Contract Nos. U-11, U-24, U-25, and the Knowledge Innovation Project
of CAS under Contract Nos. KJCX2-SW-N10, U-602, U-34 (IHEP); by the National 
Natural Science Foundation of China under Contract No. 10175060 (USTC); and by
the Department of Energy under Contract No. DE-FG03-94ER40833 (U Hawaii).


\begin {thebibliography}{99}
\bibitem{gpp} BES Collaboration, J.Z. Bai {\sl  et al.}, 
    Phys. Rev. Lett. {\bf 91}, 022001 (2003).
\bibitem{baryonium}Alakabha Datta, Patrick J. O'Donnell,
 Phys. Lett. {\bf B567}, 273(2003).
\bibitem{belle_pp1} Belle Collaboration, K. Abe {\sl  et al.}, 
 Phys. Rev. Lett. {\bf 88}, 181803 (2002).
\bibitem{belle_pp2} Belle Collaboration, K. Abe {\sl  et al.}, 
 Phys. Rev. Lett. {\bf 89}, 151802 (2002).
\bibitem{theory} See, for example, J. Ellis, Y. Frishman and M. Karlinner, 
 Phys. Lett. {\bf B566}, 201(2003);
 J.L. Rosner, Phys. Rev. D {\bf 68}, 014004 (2003);
 B.S. Zou and H.C. Chiang, Phys. Rev. D {\bf 69}, 034004 (2003).  
\bibitem{belle_pL} Belle Collaboration, M.Z. Wang {\sl  et al.}, 
 Phys. Rev. Lett. {\bf 90}, 201802 (2003).
\bibitem{BESII}BES Collaboration, J.Z. Bai {\sl  et al.},
    Nucl. Instr. Meth. A {\bf 458}, 627 (2001).
\bibitem{pdg}Particle Data Group, K. Hagiwara {\sl  et al.},
 Phys. Rev. D {\bf 66}, 010001 (2002).
\bibitem{chenjc} J.C. Chen {\sl  et al.},  
 Phys. Rev. D {\bf 62}, 034003 (2000).

\bibitem{bw} For the Breit-Wigner function, we use the form ~
$BW(M)\propto \frac{q^{2L+1} k^3 }{ {(M^2-M_0^2)}^2 + M_0^2 \Gamma^2 }$, 
where $\Gamma$ is a constant (determined from the fit), $q$ is the 
proton momentum in the $p\Lambar$ frame, $L$ is the $p\Lambar$ orbital 
angular momentum, and $k$ is the kaon momentum.

\bibitem{ssfang}  S.S. Fang {\sl  et al.}, 
 HEP \& Nucl. Phys. {\bf 27}, 277 (2003).

\bibitem{pwa} 
 For example, from the PWA fit without X(2075), we can obtain the following 
 estimation (using an acceptance about $10\%$):
 $BR(\jpsi \rightarrow p \bar{N}^*(2050)) 
  BR(\bar N^*(2050) \rightarrow K \Lambar) \sim 0.3 \times 10^{-3}$,
 $BR(\jpsi \rightarrow \Lambda \Lambar^*(1890)) 
  \sim 0.8 - 1.5 \times 10^{-3}$,
 $BR(\jpsi \rightarrow \Lambda \Lambar^*(1800)) 
  \sim 1.5 - 2.5 \times 10^{-3}$, where $N^*(2050) (J^P=3/2^+)$
  is a new resonance and the other two are listed in the PDG. 
  The above production of $\Lambda^*(1800) (J^P=1/2^-)$ is via P-wave decays, 
  and other two excited baryons can be produced via S-wave decays.
\end{thebibliography}
\end{document}